\documentclass[preprint,12pt]{elsarticle}

\usepackage{amssymb}
\usepackage{amsmath}
\usepackage{aas_macros}
\usepackage{multirow}
\usepackage{xcolor}

\journal{Physics of the Dark Universe}

\def\mn{{\mu\nu}}
\def\prt{\partial}
\newcommand{\rf}[1]{(\ref{#1})}

\begin{document}

\begin{frontmatter}



\title{Bumblebee cosmology: Tests using distance- and time-redshift probes}


\author[inst1]{Xincheng Zhu} 

\affiliation[inst1]{organization={Department of Astronomy, Tsinghua University},
            city={Beijing},
            postcode={100084}, 
            country={People's Republic of China}}
            
\author[inst1]{Rui Xu} 
\author[inst1]{Dandan Xu} 

\begin{abstract}
In modern cosmology, the discovery of the universe's accelerated expansion has significantly transformed our understanding of cosmic evolution and expansion history. The unknown properties of dark energy, the driver of this acceleration, have not only prompted extensive studies on its nature but also spurred interest in modified gravity theories that might serve as alternatives. In this paper, we adopt a bumblebee vector-tensor modified gravity theory to model the cosmic expansion history and derive predictions for the Hubble parameter. We constrain the bumblebee model parameters using observational data from established probes, including the Pantheon+ Type Ia Supernovae calibrated via the SH0ES (Supernova $H_0$ for the Equation of State) Cepheid distance ladder analysis and Baryon Acoustic Oscillations (BAO) measurements from Dark Energy Spectroscopic Instrument (DESI) Data Release 2 (DR2), as well as recently included cosmic chronometers (CC) and gamma-ray bursts (GRBs). The Markov Chain Monte Carlo (MCMC) sampling of the Bayesian posterior distribution enables us to rigorously constrain the bumblebee models and compare them with the standard $\Lambda \text{CDM}$ cosmology. We find that the bumblebee theory on its own can provide sufficiently good fits to the current observational data of distance- and time-redshift relations, suggesting its potential to explain the cosmic background dynamics. However, when compared to $\Lambda \text{CDM}$, the latter still outperforms the former according to the information criteria. We propose that further constraints from cosmological perturbation tests could impose more stringent constraints on bumblebee cosmology.
\end{abstract}




\end{frontmatter}



\section{Introduction}
\label{sec:intro}

The standard cosmological model in general relativity (GR) requires dark energy to account for the currently observed accelerated expansion of the universe \citep{1998AJ....116.1009R,2006IJMPD..15.1753C}. As the substance of dark energy is mysterious, dynamical fields are popular substitutions of dark energy \citep{CANTATA:2021ktz}. Abundant research on substituting dark energy using scalar fields has been done in the literature (for reviews see, e.g., \citep{2006IJMPD..15.1753C,2012Ap&SS.342..155B,2024CoTPh..76j7401J}). We investigate substituting dark energy using a vector field that nonminimally couples to the Ricci curvature tensor and to the Ricci curvature scalar. 
This version of vector-tensor theory of gravity is named the bumblebee gravity theory, and has been recently gaining increasing attention since the discovery of interesting black-hole solutions that pass current observational tests \citep{1973PhRvD...7.3593H,casana2018,xu2023a,xu2023b}. 
Its action is given by: 
\begin{eqnarray}
     S =&& \frac{1}{16\pi} \int d^4x \sqrt{-g} \left( R + \xi_1 B^\mu B^\nu
     R_{\mu \nu} + \xi_2 B^\mu B_\mu R \right) \nonumber
     \\&&- \int d^4x \sqrt{-g} \, \left( \frac{1}{4} B^{\mu \nu} B_{\mu \nu} + V \right),
\label{actionB}
\end{eqnarray}
where the geometrized unit system, $G=c=1$, has been used, and $g$ is the determinant of the spacetime metric $g_\mn$.
The bumblebee theory is a vector-tensor theory of gravity where the vector field $B_\mu$ modifies the behavior of gravity not only through the kinetic term $B^\mn B_\mn$ and the potential term $V$, but also through the coupling terms $B^\mu B^\nu
R_\mn$ and $B^\mu B_\mu R$. The field strength $B_\mn$ is defined in the same way as the field strength of the electromagnetic field, namely
\begin{equation}
     B_\mn = \prt_\mu B_\nu - \prt_\nu B_\mu .
\end{equation}
The two coupling constants $\xi_1$ and $\xi_2$ control the sizes of the nonminimal couplings.

Hellings and Nordtvedt first studied the theory in Equation~\rf{actionB} but without the potential term $V$ \citep{1973PhRvD...7.3593H}. Then, this theory was reintroduced by Kosteleck\'{y} with the potential term $V$ to study possible breaking of Lorentz symmetry in curved spacetime \citep{2004PhRvD..69j5009K}. To generate a vacuum expectation value for the vector field $B_\mu$ that breaks Lorentz symmetry, the potential $V$ is commonly assumed to be minimized at some nonzero configuration of $B_\mu$ \citep{2004PhRvD..69j5009K,2005PhRvD..71f5008B,2006PhRvD..74d5001B}, offering the possibility of connecting dark energy to Lorentz-symmetry violation. In this work, we intend not to dwell on the Lorentz-symmetry violating aspect of the bumblebee theory, but to focus on the simplest form of the potential,
\begin{equation}
     V = V_1 B^\mu B_\mu,
     \label{quadv}
\end{equation}
where $V_1$ is a constant. A rather interesting Friedmann-Lema\^{i}tre-Robertson-Walker (FLRW) metric solution has been found in the bumblebee theory with this potential in \cite{xuruitemp}, generalizing the FLRW solution found by Hellings and Nordtvedt \citep{1973PhRvD...7.3593H}, 
as well as by \citep{maluf21,felice2025}.
We note that a thorough study of cosmology in this theory would complete the current literature on this exemplary vector-tensor theory of gravity.
The aim of this work is to give a thorough examination of the cosmological solution using modern astronomical observations of cosmic distances and time as functions of the redshift.  

The derivation and a detailed account of the FLRW solution in the bumblebee theory 
is presented in Section 2 of \cite{xuruitemp}, with further details provided in Appendix A of the same work. We repeat the formulas here for the completeness of this paper. 
The field equations given by the action in Eq.~(\ref{actionB}) are:
\begin{align}
& G_{\mu\nu} = 8\pi T_{\mu\nu} + 8\pi \left(T_{B}\right)_{\mu\nu} + \xi_1 \left(T_{B1}\right)_{\mu\nu} + \xi_2 \left(T_{B2}\right)_{\mu\nu} , 
\nonumber \\
& D^\mu B_{\mu\nu} - \frac{dV}{dB^\nu} + \frac{\xi_1}{8\pi} B^\mu R_{\mu\nu} + \frac{\xi_2}{8\pi} B_\nu R = 0 ,
\label{fieldeqs}
\end{align}
where $T_{\mu\nu}$ consists of usual energy-momentum tensors for matter and radiation, the energy-momentum tensor for the bumblebee vector field is 
\begin{align}
\left(T_{B}\right)_{\mu\nu} =& B_{\mu\lambda}B_\nu^{\phantom\nu\lambda} - g_{\mu\nu} \left( \frac{1}{4} B^{\alpha\beta} B_{\alpha\beta} + V \right) 
+ 2 B_\mu B_\nu \frac{dV}{d(B^\lambda B_\lambda)} ,
\end{align}
and the contributions due to the couplings between the bumblebee field and the spacetime curvature are 
\begin{align}
 \left(T_{B1}\right)_{\mu\nu} =& \frac{1}{2} g_{\mu\nu} B^\alpha B^\beta R_{\alpha\beta} - B_\mu B_\lambda R_\nu^{\phantom\nu \lambda} - B_\nu B_\lambda R_\mu^{\phantom\mu \lambda} 
\nonumber \\
& + \frac{1}{2} \Big[ D_\kappa D_\mu \left( B^\kappa B_\nu \right) + D_\kappa D_\nu \left( B_\mu B^\kappa \right) 
\nonumber \\
& - \Box_g \left( B_\mu B_\nu \right) - g_{{\mu\nu}} D_\alpha D_\beta \left( B^\alpha B^\beta \right) \Big], 
\nonumber \\
 \left(T_{B2}\right)_{\mu\nu} =& -B^\lambda B_\lambda G_{\mu\nu} - B_\mu B_\nu R + D_\mu D_\nu \left( B^\lambda B_\lambda \right) 
\nonumber \\
&  - g_{\mu\nu} \Box_g \left( B^\lambda B_\lambda \right) ,
\end{align}
with $D_\mu$ being the covariant derivative and $\Box_g = D_\mu D^\mu$.

Under the cosmological principle, the universe takes the FLRW metric: 
\begin{equation}
ds^2 = a^2 \left( -d\eta^2 + \frac{dr^2}{1 - {\cal K}_0 r^2} + r^2 d\Omega^2 \right),
\end{equation} 
where $a$ is the scale factor and ${\mathcal K}_0$ represents the spatial curvature of the Universe today. Homogeneity and isotropy also require that the bumblebee field take the form of  
\begin{align}
B_\mu = \left( b_{\eta}, 0, 0, 0 \right),
\end{align}
where $b_\eta$ depends on the conformal time $\eta$. 

Considering a universe only containing ordinary matter and radiation in addition to the bumblebee field, the field equations yield two ordinary differential equations for $a$ and $b_{\eta}$:  
\begin{align}
0 =& -\frac{3 (\xi_1+2 \xi_2) a' b_\eta'b_\eta}{a} + \left( \frac{3 \xi_2 a^{\prime\,2} }{a^2} + 8\pi V_1a^2-3 {\cal K}_0 \xi_2\right) b_\eta^2  
\nonumber \\
& + 3 a^{\prime\,2} - 8\pi \left(\epsilon_{\rm m}+\epsilon_{\rm r}\right) a^4 + 3 {\cal K}_0 a^2,
\nonumber \\
0 =& b_\eta \left[ 3 (\xi_1+2 \xi_2) a'' - 3 \xi_1 \frac{a^{\prime\,2}}{a} + 6 \xi_2 {\cal K}_0 a - 16\pi V_1 a^3 \right],
\label{flrweq}
\end{align}
where $\epsilon_{\rm m}=\epsilon_{{\rm m}0}/a^3$ and $\epsilon_{\rm r}=\epsilon_{{\rm r}0}/a^4$ are the energy densities of matter and radiation, respectively; and the primes denote derivatives with respect to the conformal time $\eta$. The first equation is derived from the Einstein field equations, and the second comes from the temporal component of the vector field equation (see Eq.~\rf{fieldeqs}). 

Interestingly Eq.~\rf{flrweq} has two solutions. The first one with the bumblebee vector field vanishing (i.e., $b_{\eta}=0$), essentially reduces to the GR solution, where the universe expansion law is governed by the matter and radiation energy densities etc. The second one with the temporal component of the bumblebee field being non-zero (i.e., $b_{\eta}\neq 0$), requests the term in the big square bracket in the second equation of Eq.~\rf{flrweq} shall be zero, and essentially gives:
\begin{eqnarray}
     \frac{H(z)}{H_0} &:=& E(z) \nonumber\\
     &=& \sqrt{  \Omega_{V_1} 
          + \left( 1- \Omega_{V_1} - \Omega_{{\mathcal K}0} \right)(1+z)^{2-\alpha} 
          + \Omega_{{\mathcal K}0}(1+z)^2 },
     \label{hubblerelation}
\end{eqnarray}
where $H(z)=a^{-2}da/d\eta$ is the Hubble parameter as a function of the redshift $z$, $H_0$ is the Hubble constant, and $\Omega_{{\mathcal K}0}$ represents the current fractional density parameter corresponding to the spatial curvature of the universe. The new parameters $\Omega_{V_1}$ and $\alpha$ are related to the coupling constants $\xi_1, \,  \xi_2$ and the potential parameter $V_1$ in the bumblebee action via:
\begin{eqnarray}
&& \Omega_{V_1} := \frac{2\tilde V_1}{\xi_1+4\xi_2},
\nonumber \\
&& \alpha := - \frac{4\xi_2}{\xi_1+2\xi_2} ,
\end{eqnarray}
with $\tilde V_1 = 8\pi V_1/\left( 3H_0^2\right)$.

Equation \rf{hubblerelation} is essentially the Friedmann equation in the bumblebee universe, compared to $H(z)/H_0 = \sqrt{\Omega_{\Lambda} + \Omega_{{\rm r}0} (1+z)^4 + \Omega_{{\rm m}0} (1+z)^3 + \Omega_{{\mathcal K}0}(1+z)^2}$ for the $\Lambda$CDM universe, where $\Omega_{\Lambda}$ is the dark energy density, and $\Omega_{{\rm m}0}\equiv8\pi \epsilon_{{\rm m}0}/(3H_0^2)$ and $\Omega_{{\rm r}0}\equiv8\pi \epsilon_{{\rm m}0}/(3H_0^2)$ are the matter and radiation energy densities today, respectively. However, the former differs from the latter in the following two aspects.
(1) The role of the dark energy in the conventional $\Lambda$CDM solution is now replaced by $\Omega_{V_1}$, which is proportional to the potential parameter $V_1$, i.e., the square of the mass of the vector field. As shall be seen in later sections, this has important implication in explaining existing cosmological observations. (2) The role played by the densities of matter and radiation in the conventional $\Lambda$CDM solution (i.e., the terms of $\Omega_{{\rm r}0} (1+z)^4 + \Omega_{{\rm m}0} (1+z)^3$) is now replaced by $\left( 1- \Omega_{V_1} - \Omega_{{\mathcal K}0} \right)(1+z)^{2-\alpha}$, which rises due to the nonminimal couplings in Eq.~\rf{actionB}, resulting in a universe whose expansion is completely independent of its matter and radiation contents but solely a consequence of spacetime itself interacting with an auxiliary vector field. As already pointed out in the original paper \cite{xuruitemp}, the bumblebee cosmological model thus has a unique feature that `matter and radiation become guests visiting the prefixed background universe and cannot influence it at the homogeneous and isotropic background level', which is radically different from the $\Lambda$CDM cosmology.

We further point out that the solution in Eq.~\rf{hubblerelation} exists only when $\xi_1\ne 0$ and $\xi_2\ne0$ (for $\xi_1+4\xi_2=0$ or $\xi_1+2\xi_2=0$, Eq.~\rf{hubblerelation} is still valid by taking the corresponding limits). In fact, the bumblebee theory admits the conventional GR solution with a vanishing bumblebee field. Equation~\rf{hubblerelation} is a new solution in the bumblebee theory with a non-vanishing bumblebee field only when $\xi_1\ne 0$ and $\xi_2\ne0$. 

To investigate whether this unconventional solution offers a workable alternative to explain the universe's
late-time acceleration, various cosmological distances and time calculated using the expansion rate in Eq.~\rf{hubblerelation}, as functions of the redshift $z$, will be tested against observational datasets from Type Ia supernovae (SNe Ia) \citep{1998AJ....116.1009R, 1999ApJ...517..565P, 2022ApJ...938..110B}, baryon acoustic oscillations (BAO) \citep{1970ApJ...162..815P, 1970Ap&SS...7....3S, 2005NewAR..49..360E, 2013PhR...530...87W, 2010deot.book..246B,  2024MNRAS.534..544C, 2025arXiv250314738D}, cosmic chronometers (CC) \citep{2010JCAP...02..008S, 2016JCAP...05..014M, 2022ApJ...928L...4B, 2022LRR....25....6M, 2022JHEAp..36...27V, 2023A&A...679A..96T}, and long gamma-ray bursts (GRBs) \citep{2004ApJ...613L..13G,  2007ApJ...660...16S, 2008MNRAS.391..577A, 2022LRR....25....6M}.
In Section~\ref{sec:data}, we summarize the information of data and the structure of our Bayesian posterior inference for parameter constraints; in Section~\ref{sec:discussion}, we present the posterior inference results and compare them with the result from the conventional $\Lambda$CDM model; and in Section~\ref{sec:summary}, we make concluding remarks. The resultant constraints will be presented in terms of the parameters $H_0, \, \Omega_{{\mathcal K}0}, \, \alpha$, and $\tilde \xi_1:=\xi_1/\tilde V_1$.  

\section{Data and Analysis} \label{sec:data}
In this work, we use the observational data mentioned above to constrain the bumblebee model parameters. We use the Markov Chain Monte Carlo (MCMC) to sample the posterior distribution which is determined by both the prior information and the likelihood function. In this study, we use a Gaussian likelihood function, which incorporates a pseudo chi-square function. The likelihood is given by:
\begin{equation}
\mathcal{L} \propto \exp(-\chi^2/2),\quad\chi^2=\sum_i\frac{(y_i-y_{\text{model},i})^2}{\sigma_i^2},
\label{likelihood}
\end{equation}
where $\chi^2$ is the pseudo chi-square function, $y_i$  represent the observational data, $y_{\text{model},i}$  is the model prediction by the specific cosmological model, and  $\sigma_i$ denote the uncertainties including potential systematic errors in the observational data.


\subsection{Pantheon+ SH0ES Type Ia Supernovae}
\label{sec:SN}
Type Ia supernovae (SNe Ia) are widely used as standard candles to measure the cosmic expansion history. The luminosity distance of SNe Ia at different redshifts can be inferred using their apparent magnitudes and redshifts, given a specific cosmological model.

In this analysis, we adopt the largest SNe Ia sample published to date, the Pantheon+ compilation \citep{2022ApJ...938..110B}. It is an extended compilation of the original Pantheon analysis \citep{2018ApJ...859..101S} with 1701 SNe Ia light curves from 18 different surveys, covering a redshift range from 0.001 to 2.26. As the successor to the original Pantheon analysis, this dataset provides a more comprehensive collection of supernovae. It also includes supernovae that are in galaxies with measured Cepheid host distances and covariance (Supernova $H_0$ for the Equation of State, SH0ES analysis of Cepheid distance ladder), allowing for simultaneous constraints on parameters that describe the full expansion history of the universe. The chi-square function is defined as:
\begin{equation}
     \chi^2_{\text{SN}}=\sum_{i=1}^{1701} \frac{(\mu_{\text{obs},i}-\mu_{\text{th},i})^2}{\sigma_{\text{obs},i}^2},
\end{equation}
where $\mu_{\text{obs},i}$ is the observed distance modulus and $\sigma_{\text{obs},i}$ is the corresponding uncertainty in the observed distance modulus.
The theoretical distance modulus $\mu_{\text{th},i}$ is defined as $\mu_{\text{th},i}(z)=m-M=5\log_{10}\left(D_{\text{lum}}(z)/1\text{ Mpc}\right)+25$, where $D_{\text{lum}}$ is the luminosity distance, which is expressed as: 
\begin{equation}
D_{\text{lum}}(z) = \frac{c(1+z)}{H_0\sqrt{|\Omega_{{\mathcal K}0}|}} \cdot S_k\left(\sqrt{|\Omega_{{\mathcal K}0}|}\int^{z}_{0}\frac{dz'}{E(z')}\right),
\end{equation} 
where the function $S_k$ is defined as:
\begin{eqnarray}
S_k(x) = \left\{
  \begin{array}{ll}
    \sin(x)  & \text{ if } k=1 \\
    x        & \text{ if } k=0 \\
    \sinh(x) & \text{ if } k=-1
  \end{array}
\right.
\end{eqnarray}

\subsection{BAO measurements from DESI DR2 galaxy spectral survey}
\label{sec:BAO}
Baryon Acoustic Oscillations (BAO) trace the comoving size of the sound horizon $r_{\rm d}$ at the drag reshift $z_{\rm d}\approx 1100$ when baryons were released from photons. The BAO clustering signal can be used to measure the angular diameter distance $D_A(z)$ and the Hubble parameter $H(z)$ at different redshifts. 

We use the BAO measurements from the Dark Energy Spectroscopic Instrument (DESI) latest Data Release 2 (DR2) \citep{2025arXiv250314738D}. DESI is the most updated and advanced Stage IV galaxy survey project. The DESI DR2 features a massive dataset comprising over 14 million unique galaxy and quasar redshifts in the range $0.1<z<2.1$ based on three years of operation, and enables high-precision BAO measurements with a sophisticated blinding technique to minimize confirmation bias \citep{2025arXiv250314742A}. For the cosmology inference, these galaxy measurements are also combined with DESI Lyman-$\alpha$ forest BAO results described in \citep{2025arXiv250314739D} extending the BAO measurements to a higher effective redshift at $z=2.33$. The relation between BAO measurements and the cosmic expansion history is given by the angular diameter distance (transverse signal) and Hubble parameter (line-of-sight signal) at different redshifts. The chi-square function is defined as:
\begin{equation}
     \chi^2_{\text{BAO}}=\sum_{i=1}^{7} \frac{\left[\left( D_V/r_{\mathrm{d}} \right)_{\text{obs},i}-\left( D_V/r_{\mathrm{d}} \right)_{\text{th},i}\right]^2}{\sigma_{\text{obs},i}^2},
\end{equation}
where $D_V/r_{\mathrm{d}}$ is the ratio of the angle-averaged distance $D_V$, defined as $D_V(z)=\left[ z (1+z)^2 D_A^2(z)D_H(z) \right]^{1/3}$, to the sound horizon scale $r_{\mathrm{d}}$, with $D_A(z)$ and $D_H(z)$ being the angular diameter distance and the Hubble distance, respectively. 

\subsection{Cosmic Chronometers (CC)}
\label{sec:CC}
Cosmic chronometers (CC), as newly emerging probes, are a homogeneous class of galaxies that are passively evolving and have well-defined and synchronized star formation histories, thus serving as optimal tracers of the differential age evolution of the Universe. The age of these galaxies can be inferred from their stellar populations, which can be used to measure the cosmic expansion history. The Hubble parameter $H(z)$ can be inferred from the differential age of the universe $t(z)$ at different redshifts (see e.g., \citep{2010JCAP...02..008S, 2016JCAP...05..014M}).

The complete list of the sample we used is collectively presented in Table 1.1 of \citep{2023arXiv230709501M}. The chi-square function is defined as:
\begin{equation}
     \chi^2_{\text{CC}}=\sum_{i=1}^{35} \frac{\left( H_{\text{obs},i}-H_{\text{th},i}\right)^2}{\sigma_{\text{obs},i}^2},
\end{equation}
where $H_{\text{obs},i}$ is the observed Hubble parameter derived from $H(z)=-1/(1+z)\left({\rm d}z/{\rm d} t\right)_{\text{obs}}$, and $\sigma_{\text{obs},i}$ is the corresponding uncertainty for this measurement. Observationally ${\rm d}z$ comes from accurate spectroscopic measurements of the CC at different redshifts, while the differential age ${\rm d}t$ can be obtained by age determination through fitting stellar population model to observed spectra of the galaxies at different redshifts. 

\subsection{(Long) Gamma-Ray Bursts}
\label{sec:GRB}

Gamma-ray bursts (GRBs) are the most energetic explosions in the universe that can reach up to $10^{48}-10^{53}$ erg in a few seconds. Bright GRBs can be detected up to redshifts of 10 \citep{2022LRR....25....6M}, while the most distant detectable SN Ia is only about redshift 2.5 for now. Several correlations have been found to make GRBs (especially long bursts) as quasi-standard candles (Table 4 of \citep{2022LRR....25....6M}). The most investigated one is the Amati relation \citep{2002A&A...390...81A}, which is an empirical correlation between the peak energy of the $\gamma$-ray spectrum and the isotropic equivalent energy of the burst. 

The details of GRB samples we used can be found in \citep{PhysRevD.103.123521} and the redshift range of this dataset is from 0.0331 to 8.1. The chi-square function is defined as:
\begin{equation}
     \chi^2_{\text{GRB}}=\sum_{i=1}^{179} \frac{\left[ y_{\text{obs},i}-(kx_{\text{obs},i}+b)\right]^2}{\sigma_{\text{obs},i}^2},
\end{equation}
where $y:=\log_{10}\left(E_{\text{iso}}/1\text{ erg}\right)$, $x:=\log_{10}\left(E_p/300 \text{ keV}\right)$, with $E_p$ being the observed rest-frame spectral peak energy, and $E_{\text{iso}}=4\pi D_{\text{lum}}^2S_{\text{bolo}}/(1+z)$ representing the observed bolometric isotropic-equivalent radiated energy converted from the observable bolometric fluence $S_{\text{bolo}}$. The linear parameters $k$ and $b$ are two linear parameters, termed the Amati coefficients, describe the linear relationship between $x$ and $y$. Since
there may exist uncounted extra variability in this relation, we add an extra error term $\sigma_{\text{int}}$ as in \citep{2001ApJ...552...57R} to the total uncertainty$\sigma_{\text{obs},i}^2=\sigma_{\text{int}}^2+\sigma_y^2+b^2\sigma_x^2$.

\subsection{Joint Analysis}

Finally, we combine the likelihoods of the data into two groups: The first group includes the Pantheon+ SNe Ia data and the DESI DR2 BAO data, representing the most powerful standard cosmological probes. The other group includes all four probes, including new emerging cosmological probes, i.e., CC and GRBs. As a result, the relevant joint chi-square functions is given by:
\begin{eqnarray}
     &&\chi^2_{\text{joint,g1}}=\chi^2_{\text{SN}}+\chi^2_{\text{BAO}}, \nonumber\\&&\chi^2_{\text{joint,g2}}=\chi^2_{\text{SN}}+\chi^2_{\text{BAO}}+\chi^2_{\text{CC}}+\chi^2_{\text{GRB}}
\end{eqnarray}

\subsection{Posterior Inference}

We compare the flat and curved bumblebee models against the standard cosmology model, i.e., a spatially flat $\Lambda$CDM model, for which the only free 
parameters are the Hubble constant $H_0$ and the present-day matter density $\Omega_{{\rm m}0}$. The present-day radiation density is fixed at $\Omega_{{\rm r}0} = 9.048\times10^{-5}$ according to CMB measurements \citep{Fixsen, planck2018}.

We adopt flat priors for all model parameters: $H_0 \in [50, 100]\,[\rm {km}$ $s^{-1}\rm Mpc^{-1}]$, $\Omega_{{\rm m}0} \in [0, 1]$ (only for the $\Lambda$CDM model),  $\tilde{\xi}_1 \in [0, 2] $, $\alpha \in [-10, 2]$, and  $\Omega_{{\mathcal K}0} \in [-1, 1]$. Also we have extra priors of parameters: $\sigma_{\rm int} \in [0.1, 0.5]$, $k \in [-1, 2]$ and $b \in [52, 54]$ only for constraining Amati relation of GRB data. Using the Python package \texttt{emcee} \citep{2013PASP..125..306F}, we sample the posterior distributions with 40\,k steps. 

\begin{table}[ht]
\centering
\scriptsize               %
\caption{Marginalized constrained data of the parameters}
\label{tab:constraints}
\resizebox{\textwidth}{!}{
\begin{tabular}{lccccccc}
\hline
Model & Parameters & SNe Ia & BAO & CC & GRBs & SNe Ia + BAO & SNe Ia + BAO + CC + GRBs \\
\hline
\multirow{3}{*}{Flat bumblebee}
      & $H_0$            & $73.04\pm0.26$ & \ldots & $68.93\pm2.96$ & \ldots & $73.60\pm0.22$ & $73.60\pm0.21$ \\
      & $\tilde{\xi}_1$  & \ldots         & $0.73\pm0.02$     & \ldots         & \ldots & $0.76\pm0.02$ & $0.76\pm0.02$ \\
      & $\alpha$         & $-1.15\pm0.19$ & $-1.14\pm0.003$   & $-1.17\pm0.26$ & $-1.59\pm0.29$ & $-1.14\pm0.003$ & $-1.14\pm0.003$ \\
\hline
\multirow{4}{*}{Curved bumblebee}
      & $H_0$            & $72.81\pm0.35$ & \ldots & $66.80\pm4.00$ & \ldots & $72.89\pm0.28$ & $73.14\pm0.25$ \\
      & $\tilde{\xi}_1$  & \ldots         & $0.85\pm0.11$     & \ldots         & \ldots & $0.94\pm0.07$ & $0.85\pm0.05$ \\
      & $\alpha$         & $-0.73\pm0.32$ & $-1.18\pm0.03$    & $-0.64\pm0.38$ & $-1.70\pm0.64$ & $-1.16\pm0.006$ & $-1.15\pm0.006$ \\
      & $\Omega_{{\mathcal K}0}$ & $0.15\pm0.22$ & $0.14\pm0.09$ & $-0.03\pm0.65$ & $0.27\pm0.54$ & $0.19\pm0.05$ & $0.13\pm0.04$ \\
\hline
\multirow{2}{*}{$\Lambda$CDM}
      & $H_0$            & $72.97\pm0.26$ & \ldots & $67.75\pm3.13$ & \ldots & $73.47\pm0.20$ & $73.47\pm0.21$ \\
      & $\Omega_{{\rm m}0}$   & $0.35\pm0.02$ & $0.29\pm0.01$ & $0.33\pm0.06$ & \ldots & $0.30\pm0.01$ & $0.30\pm0.01$ \\
\hline
\end{tabular}}
\end{table}

\begin{figure*}[ht]
  \centering
  \includegraphics[width=\textwidth]{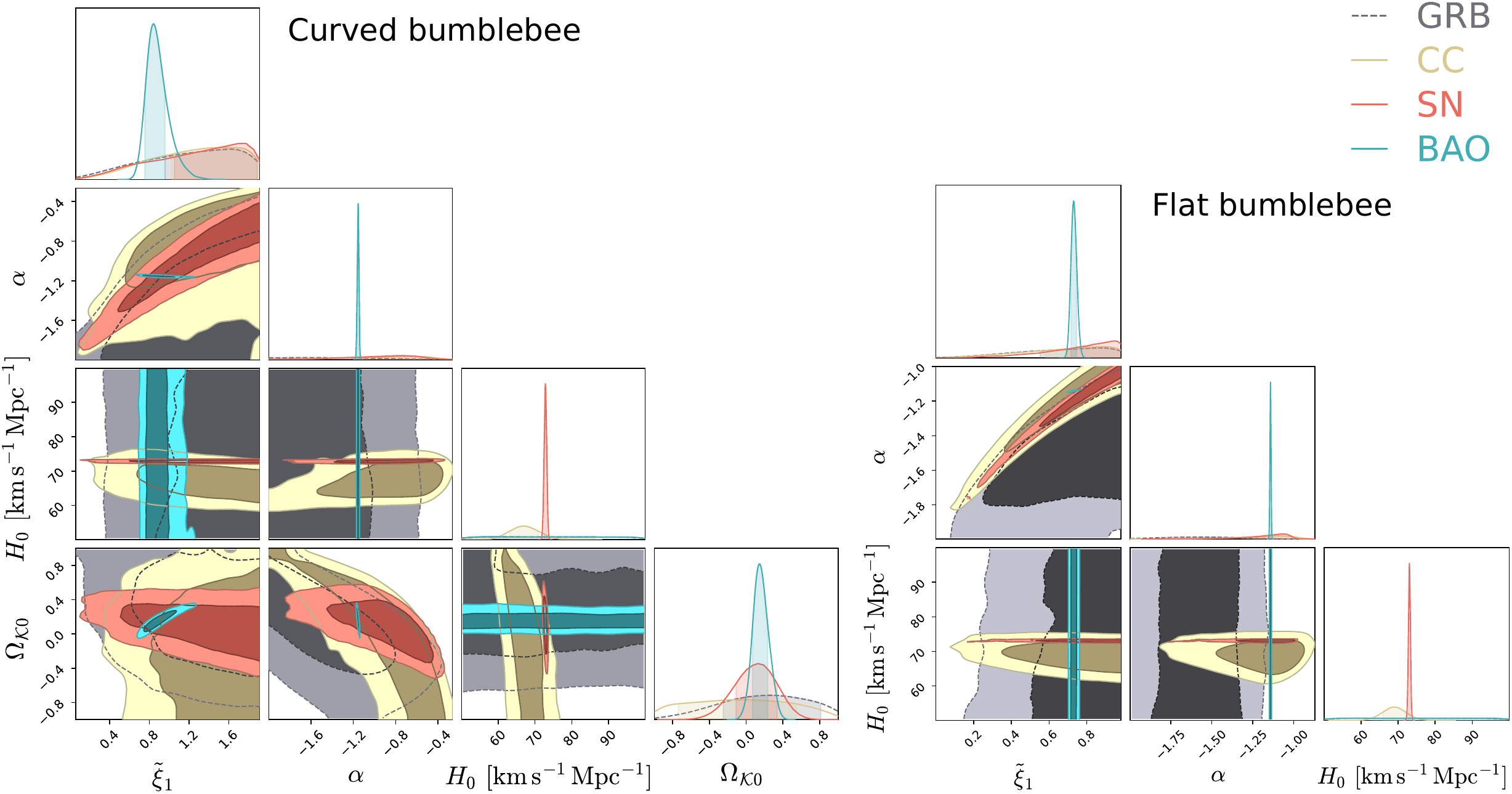}
  \caption{Posterior distributions of cosmological parameters $\tilde \xi_1$, $\alpha$, $H_0$, $\Omega_{{\mathcal K}0}$ for the spatially curved (left panel) and spatially flat ($\Omega_{{\mathcal K}0}=0$, right panel) bumblebee models, constrained by four individual cosmological probes: gamma-ray bursts (GRBs, dashed gray contours), cosmic chronometers (CC, solid yellow contours), Type Ia Supernovae (SNe Ia, solid red contours), and Baryon Acoustic Oscillations (BAO, solid cyan contours). The contours represent the 68\% and 95\% confidence regions derived from MCMC sampling.}
  \label{fig:posterior_individual}
\end{figure*}

\begin{figure*}[ht!]
  \centering
  \includegraphics[width=\textwidth]{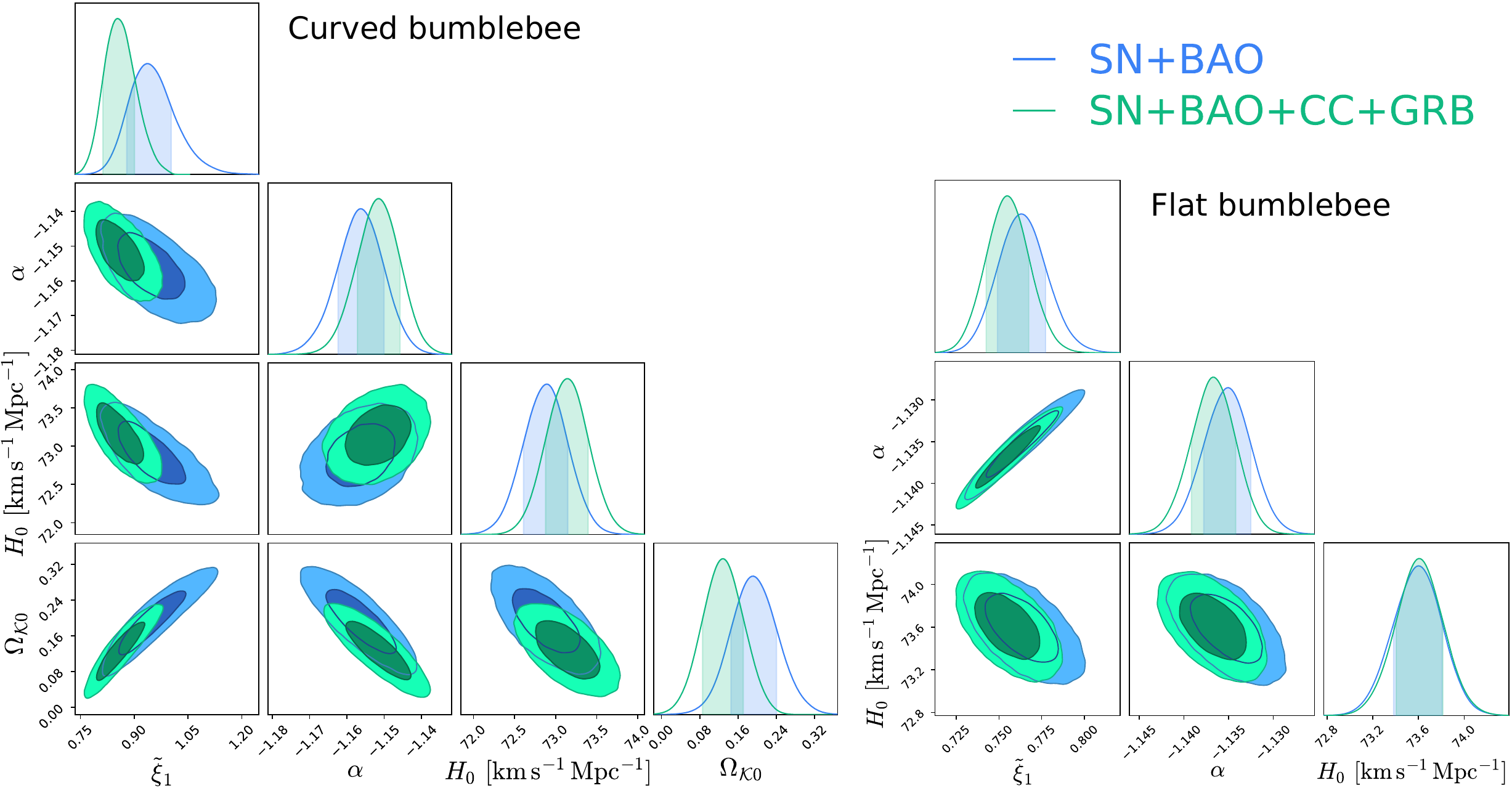}
  \caption{Posterior distributions of the same cosmological parameters as in Figure \ref{fig:posterior_individual}, for the spatially curved (left panel) and spatially flat ($\Omega_{{\mathcal K}0}=0$, right panel) bumblebee models. Results are shown for joint constraints including group 1 (blue contours), standard probes: SNe Ia + BAO and group 2 (green contours) with all 4 probes. The contours represent the 68\% and 95\% confidence regions from MCMC sampling and illustrate how combining multiple probes breaks degeneracies.}
  \label{fig:posterior}
\end{figure*}

\begin{figure*}[ht!]
  \centering
  \includegraphics[width=\textwidth]{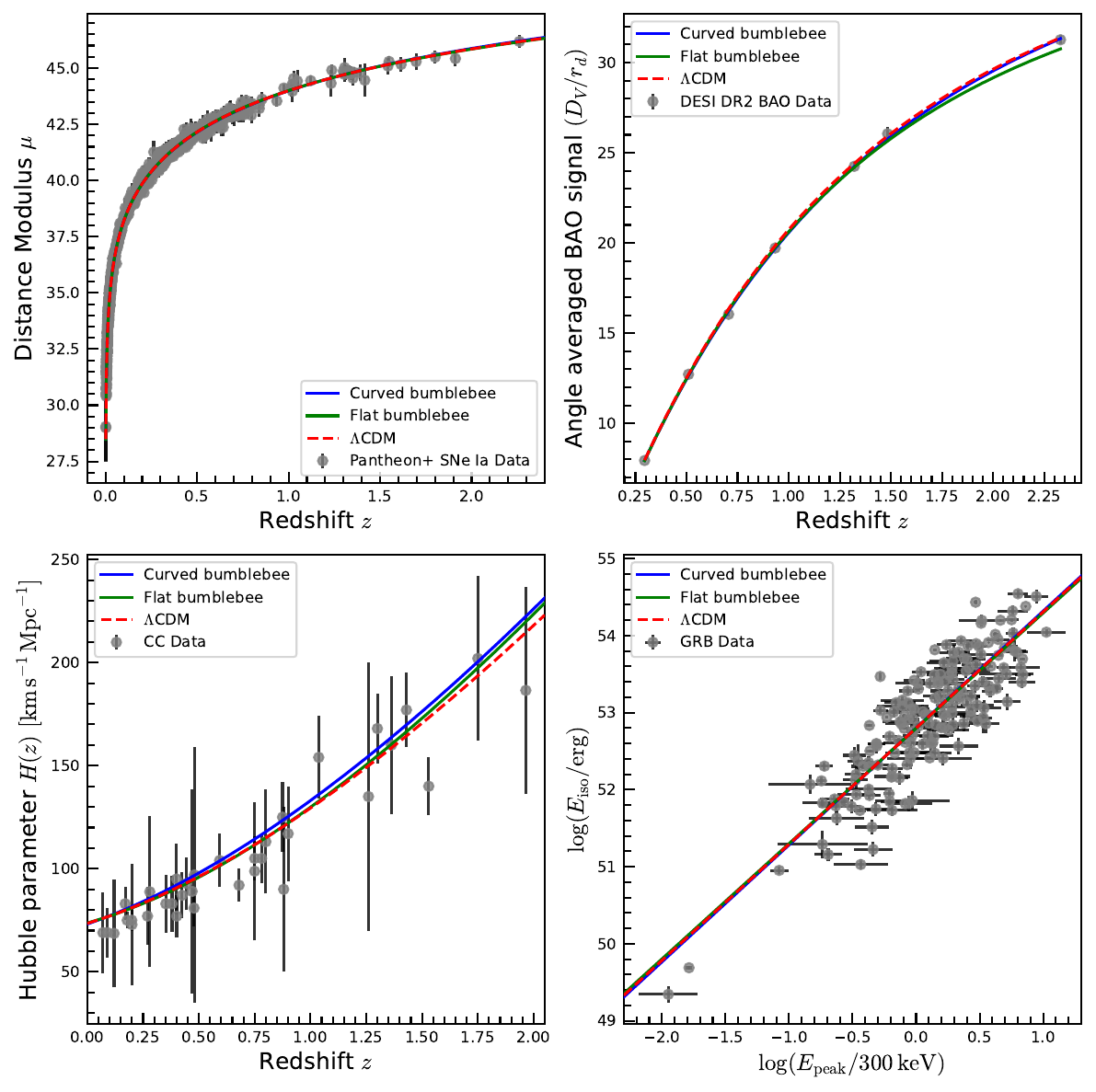}
  \caption{The best-fit prediction using all four observational probes jointly for both spatially curved and flat bumblebee model, in comparison with the standard $\Lambda$CDM model. All data points are plotted as gray filled circles with black error bars. Top left panel shows the distance modulus $\mu$ versus redshift $z$ for Pantheon+ Type Ia supernovae. Top right panel displays the angle‑averaged BAO signal $D_V/r_{\mathrm{d}}$ versus redshift $z$ from DESI DR2 measurements. Bottom left panel presents the Hubble parameter $H(z)$ versus redshift $z$ derived from CC observations. Bottom right panel depicts the Amati linear relation from GRBs. In every panel the solid blue line represents the best‑fit prediction of the curved bumblebee model, the solid green line represents the flat bumblebee model, and the red dashed line represents the $\Lambda$CDM model.}
  \label{fig:prediction}
\end{figure*}

\subsection{Information Criteria}
\label{sec:IC}

In this subsection, we compare the bumblebee model (curved or flat) with the standard $\Lambda$CDM model using the Akaike Information Criterion (AIC) and the Bayesian Information Criterion (BIC). They are statistical tools used for model selection, balancing model fit and complexity.
Introduced by \citep{1100705}, AIC is defined as:
\begin{equation}
     \text{AIC} = 2k + \frac{2k(k+1)}{N_{\text{tot}}-k-1} - 2\ln(\mathcal{L}_{\text{max}}),
\label{eq:AIC}
\end{equation}
where $k$ is the number of free parameters in the model, $N_{\text{tot}}$ denote the total sample size of the data, and $\mathcal{L}_{\text{max}}$ represents the maximum likelihood value of the model given the data. Developed by \citep{Schwarz1978EstimatingTD}, BIC also evaluates model quality but incorporates a stronger penalty for model complexity. It is calculated as:
\begin{equation}
     \text{BIC} = k\ln(N_{\text{tot}}) - 2\ln(\mathcal{L}_{\text{max}}),
    \label{eq:BIC}
\end{equation}
AIC and BIC both assess model fit while penalizing complexity, differing mainly in penalty strength. AIC applies a constant penalty per parameter, whereas BIC's penalty grows with sample size. 
From the perspective of identifying the true model, BIC appears advantageous \citep{10.1093/biomet/92.4.937}. 
In many cases, the two agree on the preferred model. When they diverge, the discrepancy itself provides valuable insights into the trade-offs between model fit and complexity. Here we have three models to compare: $\Lambda$CDM, the spatially flat bumblebee model, and the spatially curved bumblebee model. We calculate $\Delta\text{AIC}$ and $\Delta\text{BIC}$ by subtracting the AIC and BIC values, respectively, of the $\Lambda$CDM model from the corresponding values of the bumblebee models. 

\begin{table*}[ht]
\centering
\scriptsize  
\caption{Comparison of cosmological models using $\Lambda$CDM as the reference.}
\resizebox{\textwidth}{!}{
\begin{tabular}{lcccccc}
\hline
Model & $\chi^2_{\min}$ & red. $\chi^2$ &
AIC & $\Delta$AIC & BIC & $\Delta$BIC \\
\hline
\multicolumn{7}{c}{\textbf{SNe Ia + BAO}} \\
\hline
$\Lambda$CDM          & 763.81 & 0.45 & $-1138.20$ &   0.00 & $-1127.32$ &  0.00 \\
Flat bumblebee        & 766.56 & 0.45 & $-1133.45$ &   4.75 & $-1117.13$ & 10.19 \\
Curved bumblebee      & 749.57 & 0.44 & $-1148.42$ & $-10.22$ & $-1126.67$ &  0.65 \\
\hline
\multicolumn{7}{c}{\textbf{SNe Ia + BAO + CC + GRBs}} \\
\hline
$\Lambda$CDM          & 963.70 & 0.50 & $-645.79$  &   0.00 & $-618.01$  &  0.00 \\
Flat bumblebee        & 968.63 & 0.51 & $-640.93$  &   4.86 & $-607.61$  & 10.41 \\
Curved bumblebee      & 957.74 & 0.50 & $-647.88$  &  $-2.09$ & $-600.05$  &  9.00 \\
\hline
\end{tabular}%
}
\label{tab:model_comparison}
\end{table*}

\section{Results and Discussion} \label{sec:discussion}
\subsection{Constraints on model parameters using individual and joint probes}
\label{sec:modelparam}

The corresponding corner plot for posterior distributions of individual probes are presented in Figure \ref{fig:posterior_individual}. As can be seen, the SN Ia and BAO observations provide much stronger constraints than CC and GRBs, with GRBs providing the weakest constraints (and only on the bumblebee parameters $\alpha$ and $\Omega_{{\mathcal K}0}$). While SNe Ia and CC put constraints on $H_0$ but do not have sufficient constraining power on the bumblebee parameter $\tilde \xi_1$, BAO data impose strong constraints on $\tilde \xi_1$ and $\Omega_{{\mathcal K}0}$ but cannot constrain $H_0$, as the Hubble constant cancels out in the BAO measurements. 

We further show in Figure \ref{fig:posterior} the corner plot of the joint constraints using two groups of probes: (i) traditional high-precision probes, including SNe Ia and BAO; and (ii) all four probes, i.e., SNe Ia, BAO, CC and GRBs. The marginalized constraints on the bumblebee parameters are summarized in Table \ref{tab:constraints}. The fact that the overlapping regions of the $2\sigma$ posterior distributions for individual probes align with the central $2\sigma$ region of the joint distribution indicates that the four probes are consistent for both flat and curved bumblebee models and there is no statistical tension between the probes. 

Comparing the results using only traditional high-precision probes (SNe Ia and BAO) with those incorporating the emerging probes CC and GRBs, we observe a slight shift in the center of the posterior distribution. The limited impact of the emerging probes can be attributed to two factors. First, as shown in Figure \ref{fig:posterior_individual}, the posterior distribution of CC is similar in shape to that of SNe Ia, albeit with a wider range due to larger systematic errors. The relatively larger errors and fewer data points compared to SNe Ia result in a smaller contribution to the total $\chi^2$. Second, GRB measurements suffer from significant uncertainties. While the Amati relation suggests a well-aligned linear relationship, the large intrinsic dispersion, substantial measurement errors, and the small sample size (179 GRBs) greatly weaken the constraining power of GRBs as a probe (Figure \ref{fig:posterior_individual}). Although these two new probes contribute only marginally to the results, we argue that combining multiple probes is still of great significance. On one hand, emerging probes offer independent consistency checks, which can test the internal coherence of gravity models. On the other hand, with continuous advancements in observational techniques, the precision of these probes is expected to improve significantly in the foreseeable future \citep{2022LRR....25....6M}. Over time, their advantages may become more evident: the constraints from CC on $H(z)$ come from differential measurements on galaxy ages, and thus are independent of the (cosmology-model dependent) absolute galaxy ages. Meanwhile, GRBs probes extend to significantly higher redshifts compared to traditional probes, with minimal sensitivity to dust or gas absorption. The inclusion of these complementary probes represents a forward-looking strategy with substantial potential to advance cosmological studies. 

Figure \ref{fig:prediction} presents all four observational data sets individually, over-plotted with the best model predictions jointly constrained by all four data sets for the two bumblebee models---one allowing spatial curvature and the other assuming a flat universe---and the $\Lambda$CDM model. We note that the best-fit predictions to each individual probe are made using maximum a posteriori (MAP) point parameters of the joint posterior distribution that is constrained using all four probes together. 
As can be seen, for all three models, the best model predictions in general provide a good fit to all four probes.

The reduced $\chi^2$ values for different models are presented in Table \ref{tab:model_comparison}. With a reduced $\chi^2$ value of $0.4-0.5$, the bumblebee models provide sufficiently good fits to the data. In the case of the curved bumblebee model in particular, the results show that all four probes jointly favor an open bumblebee universe with a curvature parameter $\Omega_{{\mathcal K}0} \sim 0.13$. It is worth noting that to date there is no strong and direct evidence suggesting a non-trivial topology for our universe. While standard inflation predicts $\Omega_{{\mathcal K}0} = 0$, modified inflationary scenarios like ``open inflation'' allow for $\Omega_{{\mathcal K}0} >0$. In the open inflation model, a significantly sub-unity density parameter is naturally explained, supporting the possibility of a non-flat universe (\citep{1995PhRvD..52.5538B, 1998PhRvD..58h3514L}). 

The fitting results suggest that $\Omega_{V_1} \sim 72\%$ and 63\% in the flat and curved bumblebee models, respectively. Comparing to the $\Lambda\text{CDM}$ model, the bumblebee parameter $\Omega_{V_1}$ can be interpreted as the equivalent of dark energy, which is estimated to be $\sim 70\%$ in $\Lambda\text{CDM}$. The evolution index ($2-\alpha$) for the term $1-\Omega_{V_1}-\Omega_{{\mathcal K}0}$ (see Eq.~\rf{hubblerelation}) is constrained to be 3.14 and 3.15 for the flat and curved bumblebee universes, respectively. In comparison, this term is equivalent to the sum of the matter and radiation densities in $\Lambda\text{CDM}$, which evolve with a slope of 3 and 4, respectively. It is interesting to notice the closeness in these model parameters between the bumblebee cosmology and $\Lambda$CDM. We attribute the good performance of the bumblebee models to its close resemblance in their behavior to $\Lambda$CDM.

As a side remark, we note that cosmological inference always presents correlations among model parameters. These correlation sometimes reflect real physical constraints required by the model, such as $\Omega_{{\rm m}0} + \Omega_{\Lambda} = 1$ for a flat universe. At other times, they arise because multiple parameters jointly determine the same observable quantity and may appear degenerate when observational constraints are limited. In the bumblebee models, the strong correlations appearing between parameter pairs such as $\Omega_{\mathcal{K}0}$ and $\xi_1$, and $\alpha$ and $\xi_1$, are more likely due to parameter degeneracies rather than physical constraints. This can be seen from Figure \ref{fig:posterior_individual}, as different probes actually show different covariance distributions among these parameter pairs.

\subsection{Model comparison: Bumblebee vs. $\Lambda\text{CDM}$}
\label{sec:modelcompare}

When comparing models, one should use information criteria to assess which model is better supported by the data. According to Jeffreys's scale \citep{kass1995bayes}, the condition $\Delta\text{IC} < 2$ confirms the statistical compatibility of the two models. A range of $2 < \Delta\text{IC} < 6$ suggests mild tension, while $\Delta\text{IC} \geq 10$ implies strong disagreement between the models.
Table \ref{tab:model_comparison} shows the corresponding minimum $\chi^2$ and the reduced $\chi^2$ values, together with information criteria (ICs) among different models constrained by either traditional high-precision probes alone (SNe Ia and BAO) or further combined with emerging probes of CC and GRBs. 
In each case the AIC and BIC values for each model are calculated by the best-fit parameters of the joint all four probes.
As can be seen, with a reduced $\chi^2$ of 0.45 and 0.5, the two bumblebee models (flat and curved) on their own can indeed provide sufficiently good fit to the current observational data of distance- and time-redshift relations shown in Figure \ref{fig:prediction}. This suggests the potential of the bumblebee cosmological models in explaining the cosmic background dynamics. 

When comparing to the $\Lambda\text{CDM}$ model, the flat bumblebee model shows mild tension with $\Lambda\text{CDM}$ in AIC ($\Delta\text{AIC} \sim 4.8$) and even receives strong opposition from BIC ($\Delta\text{BIC} \sim 10$) due to higher complexity penalties in BIC. In contrast, the curved bumblebee model significantly outperforms $\Lambda\text{CDM}$ under traditional constraints (SNe Ia + BAO) ($\Delta\text{AIC} \sim -10$) while remaining statistically indistinguishable in BIC ($\Delta\text{BIC} \sim 0.6$). When combining all four probes, the curved bumblebee model maintains statistical compatibility with $\Lambda\text{CDM}$ in AIC ($\Delta\text{AIC} \sim -2$) but becomes strongly disfavored by BIC evaluation ($\Delta\text{BIC} \sim 9$).

It is worth noting that the joint constraints have revealed a divergence in the conclusions drawn from AIC and BIC: while AIC favors the curved bumblebee model, BIC prefers the $\Lambda\text{CDM}$ model. This discrepancy arises from the different penalties applied to model complexity $k$. 
Both AIC and BIC as considered in this study have their merits and limitations depending on the context. As can be seen from Table \ref{tab:model_comparison}, $\Delta$BIC is consistently larger than $\Delta$AIC, indicating that BIC penalizes the bumblebee models --- being more complex than $\Lambda$CDM ---more heavily than AIC. The seemingly contradictory conclusions from AIC and BIC when comparing the curved bumblebee model to $\Lambda$CDM --- particularly in the case of all four probes being jointly used---can be attributed to the fundamental differences between the two criteria. As can also be seen from Eqs.~\rf{eq:AIC} and \rf{eq:BIC}, AIC tends to favor more complex models (e.g., the curved bumblebee model), while BIC imposes a stronger penalty on complexity and thus prefers simpler models, especially when the dataset is large (e.g., $\Lambda$CDM under joint constraints from all four probes).
Here we will not further debate the superiority between AIC and BIC, instead we present both results to provide comprehensive insights for future research.

\section{Conclusions and Summary} \label{sec:summary}
In this paper, we present Bayesian posterior probability constraints on an unconventional cosmological model based on the bumblebee gravity theory using multiple cosmological probes. Our approach has incorporated emerging probes such as cosmic chronometers (CC, Section~\ref{sec:CC}) and gamma-ray bursts (GRBs, Section~\ref{sec:GRB}) alongside well-established SNe Ia (Section~\ref{sec:SN}) and BAO datasets (Section~\ref{sec:BAO}). 
We show that the constraints from the four probes are consistent within the $2\sigma$ level, and the best-fit results from the bumblebee cosmology under the joint constraints can effectively explain each individual and the combined dataset (see Figure \ref{fig:prediction} and Table \ref{tab:model_comparison}). In particular, the curved bumblebee model favors an open universe with $\Omega_{{\mathcal K}0} \sim 0.13\pm 0.04$, challenging the traditional flat universe geometry (see Section~\ref{sec:modelparam}).  

When compared to the standard $\Lambda\text{CDM}$ framework, information criterion calculations (AIC and BIC, see Section~\ref{sec:IC}) indicate a general superiority of $\Lambda\text{CDM}$ over the bumblebee models, except for the curved bumblebee model when evaluating using AIC (see Table \ref{tab:model_comparison} and Section~\ref{sec:modelcompare}). It is worth noting that our results demonstrate a sufficiently good performance of the bumblebee theory in fitting the distance- and time-redshift relations, and thus its potential in explaining the cosmic background dynamics. 
However, we remind the reader that cosmological tests of the theory in fact lie in two aspects. The first category of constraints comes from cosmic background dynamics, i.e.,  through observational imprints of the cosmic acceleration (as investigated in this paper). The second category is at cosmological perturbation level, e.g., through the cosmic microwave background (which is carried out in a parallel study). It is important to not only use modern distance- and time-redshift probes, but also utilize constraints from the cosmological perturbation perspective,
in order to provide more stringent constraints on the bumblebee cosmology as well as other modified gravity theories (see a parallel study from \citep{xuruitemp}).

Finally we mention in passing that the recent development and incorporation of emerging cosmological probes offer opportunities to perform independent consistency checks on models (as here aiding in the assessment of whether a modified gravity theory remains viable and in achieving tighter parameter constraints). However, we must emphasize that before combining new probes for joint posterior analysis, it is critical to verify that no significant biases exist between them and established probes. Severe biases could create tension between probes under the same model, rendering joint constraints meaningless and potentially leading to incorrect rejection of the model. Additionally, new methods should carefully evaluate potential sources of systematic error, adopting the most conservative error estimates to avoid prematurely discarding valid models.


\section*{Acknowledgements}
We thank Prof. Cheng Zhao and an anonymous referee for valuable comments. This work was supported by the National Natural Science Foundation of China (Grants No. 12405070), the China Postdoctoral Science Foundation (2023M741999, GZC20240872), and the high-performance computing cluster in the Astronomy Department at Tsinghua University.





\bibliographystyle{elsarticle-num} 
\bibliography{els}






\end{document}